\documentclass[aps,prl,twocolumn,superscriptaddress,showpacs]{revtex4}

\usepackage{longtable}
\usepackage{graphicx}
\usepackage{dcolumn}
\usepackage{bm}
\usepackage{amsmath}
\usepackage[psamsfonts]{amssymb}

\newcommand{\STO}{SrTiO$_3$}
\newcommand{\LAO}{LaAlO$_3$}

\newcommand{\etal}{\emph{et al.}}

\newcommand{\Hperp}{H$_{\perp}$}
\newcommand{\Hcperp}{H$_{c2\perp}$}
\newcommand{\Hpar}{H$_\parallel$}
\newcommand{\Hcpar}{H$_{c2\parallel}$}
\newcommand{\Hstar}{H$^*$}
\newcommand{\Eso}{$\epsilon_{SO}$}

\begin{document}

\title{Tuning spin-orbit coupling and superconductivity at the \STO/\LAO~ interface:\\ a magneto-transport study}


\affiliation{Raymond and Beverly Sackler School of Physics
and Astronomy, Tel-Aviv University, Tel Aviv, 69978, Israel}

\author{M. Ben Shalom}
\author{M. Sachs}
\author{D. Rakhmilevitch}
\author{A. Palevski}
\author{Y. Dagan}
\email[]{yodagan@post.tau.ac.il}


\date{\today}

\begin{abstract}
The superconducting transition temperature, T$_c$, of the
\STO/\LAO~ interface was varied by the electric field effect. The
anisotropy of the upper critical field and the normal state
magneto-transport were studied as a function of gate voltage. The
spin-orbit coupling energy $\epsilon_{SO}$ is extracted. This
tunable energy scale is used to explain the strong gate dependence
of the mobility and of the anomalous Hall signal observed.
$\epsilon_{SO}$ follows T$_c$  for the electric field range
under study.

\end{abstract}


\maketitle
Interfaces between strongly correlated oxides exhibit a variety of
physical phenomena and are currently at the focus of intensive
scientific research. An electronic reconstruction occurring at the
interfaces may be at the origin of these
phenomena.\cite{OkamotoMillis} It has been demonstrated that the
interface between \STO~(STO) and \LAO~(LAO) is highly conducting,
having the properties of a two dimensional electron (2DEG) gas.
\cite{OhtomoHwang} At low temperatures the 2DEG has a
superconducting ground state, whose critical temperature can be
modified by an electric field effect.\cite{N.Reyren08312007} The
origin of the charge carriers and the thickness of the conducting
layers are still under debate. \cite{Nakagawa_no_O_defects,
pentchevaPicketPRL, popovictheoryfortwodeg, PhysRevLett.98.196802,
WillmottPRL, Basletic_thickness, OkamotoMillis} Recently, we have
demonstrated that for carrier concentrations at the range of 10$^{13}$
cm$^{-2}$, a large, highly anisotropic magnetoresistance (MR)
is observed.\cite{benshalom} Its strong anisotropy suggests that
it stems from a strong magnetic scattering confined to the
interface.
\par
Here we show that both superconducting and spin-orbit (SO)
interaction can be modified by applying a gate voltage. The upper
critical field applied parallel to the interface, \Hcpar, is
approximately the weak coupling Clogston-Chandrasekhar
paramagnetic limit: $\mu_0H_{c2}\cong1.75k_BT_c $ for high carrier
concentrations. However, as the density of the charge carrier is
reduced, \Hcpar~becomes as large as five times this limit,
suggesting a rapidly increasing SO coupling. This SO coupling energy
$\epsilon_{SO}$ manifests itself in many of the transport
properties studied.
\par

\begin{figure}
\includegraphics[width=1\hsize]{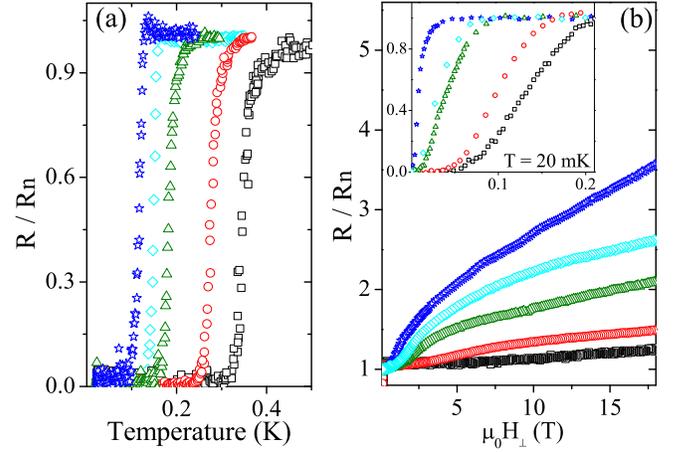}
\caption {Color on-line (a) Normalized sheet resistance $versus$
temperature for the various gate voltages (from left to right)
V$_g$=50 V, as-grown state, 10 V, -10 V, -50 V color and shape
code applies to all graphs in the paper. (b) Resistivity $versus$
the perpendicular magnetic field . Inset:  zoom on the low field
region where superconductivity is suppressed.\label{Hc_Tc}}
\end{figure}

\par 
We use a sample with 15 unit cells of \LAO, deposited by pulsed
laser on an atomically flat  \STO(100) substrate. The oxygen pressure during the
deposition was maintained at $10^{-4}$ Torr. The deposition was
followed by a two-hour annealing stage at oxygen pressure of 0.2
Torr and a temperature of 400 C. The thickness of the \LAO~layer was monitored by
reflection high energy electron diffraction. Growth procedure is
similar to the one described elsewhere.\cite{benshalom} A layer of
gold was evaporated at the bottom of the sample and used as  a gate
when biased to $\pm$50 V relative to the 2DEG. Contacts were made
using a wire bonder in a Van Der Pauw geometry. We took extra care
to ensure the absence of heating by testing the resistance to be
current-independent at the superconducting transition point, where
current sensitivity is maximal. During measurements, the sample was cooled
down to a base temperature of 20 mK and a magnetic field of up to
18 T was applied. Samples rotation was done using a step motor
with a resolution of 0.015$^{\circ}$/step. All transport
properties reported here were measured in the as-grown state
(AG), and while applying V$_g$=50, 10, -10, -50 V. In order to
accurately determine T$_c$, the remnant magnetic field was
minimized by oscillating the field down to zero. In addition, the
sample was kept parallel to the field where superconductivity is
less sensitive to it.
\par

Fig.$\ref{Hc_Tc}$a presents the normalized resistance R/R$_n$
as a function of temperature for the various V$_g$ values. R$_n$ is the resistance measured at zero
field and at T=0.5 K, well above T$_c$, for each V$_g$. Tuning
V$_g$ from +50 to -50 V increases T$_c$ from 0.1 to 0.35 K.
\par
Fig.\ref{Hc_Tc}b presents the resistance normalized with R$_n$ $versus$ the perpendicular magnetic field at T=20 mK. A large,
positive MR is observed for V$_g$=50 V. Its magnitude gradually
decreases as V$_g$ is reduced. The inset zooms on the
superconducting low field region.
\par
Table 1 summarizes the superconducting properties for the various V$_g$.
We define \Hcperp~ (T$_c$) as
the field (temperature) at which the resistance is R$_n$/2. The Ginzburg Landau coherence length
$\xi_{GL}=\sqrt{\Phi_{0}/2\pi\mu_{0}H_{c \perp}}$ is extracted from
\Hcperp, here $\Phi_{0}$ is the flux quantum.
\par
The charge carrier density $n=1/R_He$ is inferred from Hall
measurements in the high field regime ($\mu_0H>$14 T). It is
presented together with the calculated mobility $\mu=n/R_n$.
A large variation of $\mu$ is observed,
indicating a significant change in the scattering rate with gate
voltage as previously noted.\cite{Hwangmobility} This behavior
will be discussed later.

\begin{figure}
\includegraphics[width=1\hsize]{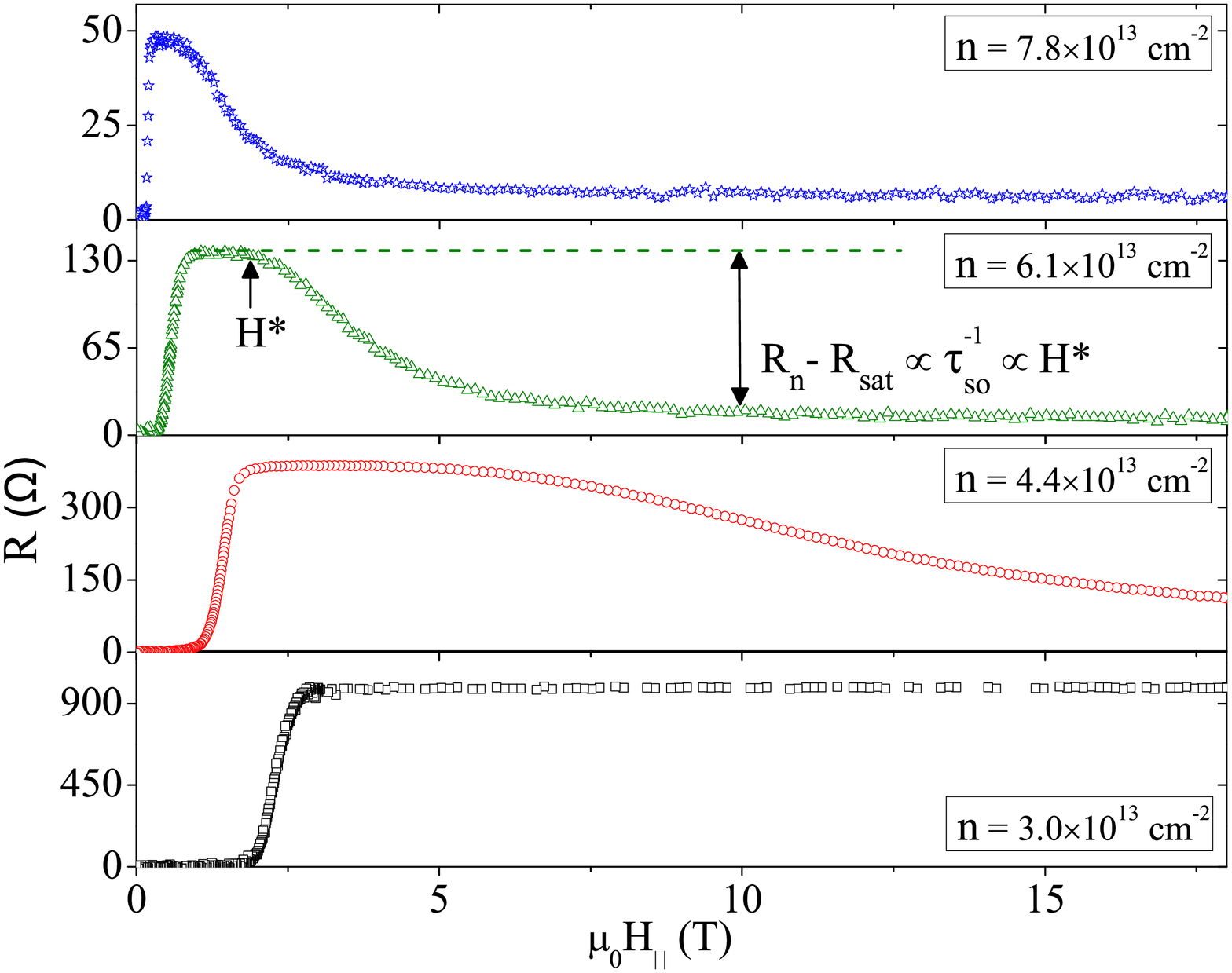}
\caption {Sheet resistance $versus$ the magnetic field applied
parallel to the interface and current for various charge
densities.\label{H_parallel}}
\end{figure}

Fig.\ref{H_parallel} presents the sheet resistance $versus$
the magnetic field, \Hpar, applied parallel to the interface and to
 the current. From the
low field regime we extract the superconducting parallel critical
field \Hcpar=H(R$_n$/2). We note that while T$_c$ increases merely
to 0.35 K, \Hcpar~ becomes as large as 2.5 T [table 1]. The high \Hcpar~ and
low T$_c$ implies that the paramagnetic limit
 is exceeded.\cite{Clogston} Since $\xi_{GL}$ is rather large, 50-200 nm, it is reasonable to use the
weak coupling BCS approximation. Accordingly, we expect
superconductivity to exist in fields lower than $g\mu_{B}H\leq
3.5k_{B}T_c$; here g is the gyromagnetic ratio and $\mu_B$ is the
Bohr magneton.
\par
\begin{figure}
\includegraphics[width=1\hsize, height=0.75\hsize]{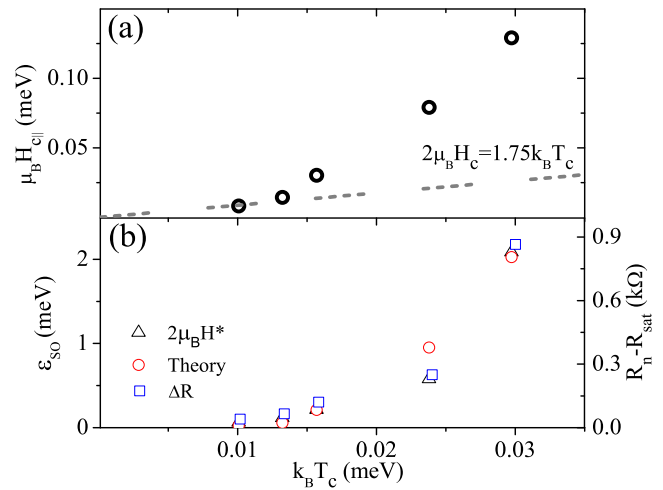}
\caption {(Color on-line) a. The parallel upper critical field
H$_{c\parallel}$ $versus$ k$_B$T$_c$ while V$_g$ is varied. The
dashed line represents the expected behavior for the
paramagnetic limit.\\ b. $\epsilon_{SO}$ extracted from \Hstar~ [see Fig.\ref{H_parallel}], and  calculated from the measured superconducting properties (\Hcpar~and T$_c$),\cite{PhysRevB.12.877}  $versus$ k$_B$T$_c$. The right axes presents $\Delta$R, which scales with $\epsilon_{SO}$ as expected. \label{Pauli}}
\end{figure}

In Fig.\ref{Pauli}a $\mu_{B}$\Hcpar~is plotted against $k_{B}T_c$
for different V$_g$ values. The dashed straight line represents
the paramagnetic limit using BCS weak coupling and g=2. As T$_c$
increases, the limit is exceeded by up to a factor of $\sim$ 5.
This behavior suggests a strong SO coupling that relaxes the Clogston-Chandrasekhar
limitations.\cite{Clogston}
For a superconducting layer thickness $d\ll \xi_{GL}$ it is
expected that \Hcpar~be much larger than \Hcperp.  In
this case the critical field is determined by the paramagnetic  limit and
the spin-orbit energy. From Fig.\ref{Pauli}a it appears that even
for the lowest T$_c$, superconductivity is quenched in this
paramagnetic limit. We can therefore merely set an upper limit on
the thickness of the conducting layer by analyzing the anisotropy
of the critical field:
$d\leq\sqrt{3}\Phi_{0}\mathbf{/}\pi\xi_{GL}\mu_{0}$\Hcpar. \cite{Guy} The
thickness upper limit is presented
 for the various V$_g$ reaching a minimal value of 10 nm
for $n=3\times10^{13} cm^{-2}$ [table 1]. Previous estimations obtained
similar values. They, however, relate $d$ to the actual thickness.\cite{Reyrend, Hwangnature}
\par

Let us describe the general behavior of R(\Hpar)
[Fig.\ref{H_parallel}] using the $n=6.1\times 10^{13} cm^{-2}$
curve (green triangle) as an example: Above the superconducting transition, the
sheet resistance reaches a roughly field-independent regime
(0.9-1.8 T) in which the resistance approximately equals R$_n$. We
define \Hstar~as the onset field where $dR/dH$ becomes negative. For H$>$\Hstar~the resistance drops to R$_{sat}$, the
low resistance saturation regime (\Hpar$>$10 T). Both \Hcpar~ and
\Hstar~ strongly depend on V$_g$; they increase as V$_g$ changes
from 50 to -50 V, and \Hstar~ becomes unmeasurably high for
V$_g$=-50 V (black squares).
\par

\begin{table}[ht]
\caption{} 
\centering      
\begin{tabular}{c c c c c c c c c c}  
\hline\hline                        
Vg  & n$\times10^{13}$ & Mobility  &  $Tc$  & $\mu_0Hc_{\perp}$  &  $\mu_0Hc_{\|}$  & d & $\xi_{GL}$ & $\epsilon_{SO}$  \\ [0.5ex] 
V & $cm^{-2}$ & $cm/sec$V  & K & T & T & nm & nm & meV \\ [0.5ex]
\hline                    
-50 & 3.0 & 236   & 0.35 & 0.125 & 2.23 & 10   & 51 &$>$2 \\   
-10 & 4.4 & 392   & 0.28 & 0.098 & 1.37 & 14.4 & 58 & 0.58 \\
10  & 6.1  & 762   & 0.18 & 0.040 & 0.52 & 24.1 & 90 & 0.21 \\
50  & 7.8  & 1707 & 0.12 & 0.009 & 0.14 & 41.2 & 194 & 0.06\\
AG & 9.5  & 897    & 0.15 & 0.030 & 0.25 & 44    & 103 & 0.12\\ [1ex]       
\hline     
\end{tabular}
\label{table:nonlin}  
\end{table}

Fig.\ref{Pauli}a suggests that SO coupling plays a major role in the system. We shall now describe the data presented so far using a single energy scale, \Eso. We relate \Hstar~  to the breakdown field of the spin-orbit coupling
energy: $g\mu_B$H$^*=\epsilon_{SO}$. The
$\epsilon_{SO}$ values are presented assuming g=2 [table 1]. Above this field SO scattering
is suppressed and completely vanishes at the high field
saturation regime where all spins are aligned. In this scenario
the high field saturation value R$_{sat}$ is the remnant impurity
scattering. The SO scattering rate can be evaluated from the
difference between R$_n$ and R$_{sat}$. Therefore R$_n$-R$_{sat}$
should be proportional to $h/\tau_{SO}=\epsilon_{SO}$.
\par
Fig.\ref{Pauli}b presents R$_n-R_{sat}$ and $\mu_B$\Hstar~$versus$
$k_BT_c$. These two quantities scale as predicted. Furthermore,
the saturation resistance R$_{sat}$ roughly scales with the number of
carriers.
We further use the calculation of R. A. Klemm \etal~ which estimates
the SO scattering time from \Hcpar~and T$_c$ for a two-dimensional film in the paramagnetic limit.\cite{SpinorbitHcpar}
As seen [Fig.\ref{Pauli}b, red circles], the general behavior is described by this (probably oversimplified) model.
We can therefore conclude that a single energy scale, tuned
by V$_g$, dominates
the behavior of the transport and affects superconductivity.
For the carrier concentration range under study R$_{sat}\ll R_{n}$.
Hence, the main contribution to the zero field resistance comes from SO
scattering. This explains the strong dependence of the mobility on
gate voltage.

\par
We note that for V$_g=50$ V the carrier concentration is lower compared to the as-grown state. This is in contrast to a simple capacitor-like behavior with negative charge carriers.
Furthermore, despite the decrease in carrier concentration, R$_n$ unexpectedly decreases. We relate this peculiar behavior to the SO scattering processes dominating R$_n$. We assume that static positive charges move to the interface
when the highest positive voltage is applied and consequently $n$ is reduced.
This charge movement reduces the initial electric field at the interface and as a result decreases the SO scattering.
Upon reducing the gate voltage from its maximal positive value, electrons are drawn away from the interface in a
reversible manner as expected, and an electric field is created.
The initial electric field in the as-grown state results in enhanced SO scattering and higher resistance, despite the higher carrier concentration.
\par

\begin{figure}
\includegraphics[width=1\hsize]{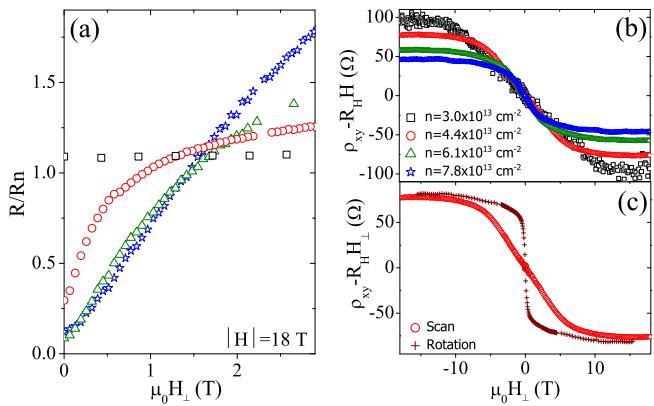}
\caption {(Color on-line) a. Normalized
resistance as a function of perpendicular magnetic field. Data is
taken while rotating the sample in a constant magnetic field of 18 T. b. Hall resistivity after subtraction of a linear fit for high fields (AHE). c. The AHE for $n=4.4\times10^{13}cm^{-2}$ while H is scanned, and while rotating the sample at a constant field.  \label{Rotation}}
\end{figure}

The MR presented [Fig.\ref{Hc_Tc}b, Fig.\ref{H_parallel}] is
highly anisotropic for high magnetic fields. For example, for
$n=6.1\times10^{13}cm^{-2}$ and H=18 T, the resistance vary from 0.12R$_n$ to 3.5R$_n$
(a factor of 30) for the in-plane and out-of-plane field configurations respectively.
We study this anisotropy by rotating the sample around an in-plane axis which is perpendicular to the current, while keeping the total field constant,
$|\vec{H}|=$18 T.
Fig.\ref{Rotation}a presents the normalized
sheet resistance $versus$ the perpendicular field
component. For the small angle range presented, the parallel field
component is approximately constant $\simeq$18 T. Moreover, for
this parallel field component, the resistance is insensitive to
small changes in \Hpar [Fig.\ref{H_parallel}]. As shown, the sheet
resistance rapidly increases when a small perpendicular component
is introduced. It reaches R$_n$ for \Hperp$\simeq$1.5 T (an angle
of about $4^\circ$).
 This is similar to our previous observation at 2 K.\cite{benshalom}
The curves $n=7.8, 6.1\times10^{13}cm^{-2}$ and AG (not shown) merge near 1.5 T, while $n=4.4\times10^{13}cm^{-2}$
departs form the general behavior. This can be easily explained since
for this gate voltage the resistance does not saturate up to 18 T
[Fig.\ref{H_parallel}]. For $n=3.0\times10^{13}cm^{-2}$, H=18 T is well below
\Hstar~and therefore the anisotropy cannot be observed at this too
small a magnetic field. We relate the sudden appearance of resistance
with a small perpendicular field component to the orbital motion generated
by this component. This motion is enhanced by the high mobility state (R$\simeq$R$_{sat}$) induced by the presence of a large parallel magnetic field.
\par
Fig.\ref{Rotation}b presents the Hall resistivity $\rho_{xy}$ at
20 mK $versus$ the magnetic field
after subtracting out the linear term obtained from a fit to the
high field regime. A conspicuous deviation from this linear part
in the Hall resistivity (anomalous Hall effect AHE), is
observed. This AHE persists up to 100 K with a peculiar, roughly linear
temperature dependence (not shown). As shown, the applied
V$_g$ varies the AHE saturation field. This variation is inconsistent with a
simple magnetic impurity scenario and with a simple ferromagnetic behavior.
Furthermore, no hysteresis is observed for all carrier concentrations.
Here we note that the AHE saturation
field roughly follows the behavior of $\epsilon_{SO}$ and may be
related to it.

\par
Fig.\ref{Rotation}c compares the
data in Fig.\ref{Rotation}b in the case where $n=4.4\times10^{13}cm^{-2}$
(red circles), with data taken at constant magnetic field while rotating the sample (as in Fig.\ref{Rotation}a). For the rotation (brown pluses), the slope of the AHE is much steeper. The two measurements differ at low
perpendicular fields where the mobility is an order of magnitude
higher for the rotation. The qualitative variation between the measurements is expected since the AHE is
usually proportional to some power of the resistivity. The AHE has
been related to a multiple band structure (light and heavy)\cite{Hwangmobility} and to incipient disordered magnetism induced by the field\cite{seri:180410}. The former is unlikely in view of the strong AHE
slope in the case of the rotation experiment. For the case of light and heavy
charge carriers, one would expect the former to dominate the
resistivity and the latter to dominate the low field Hall.
However, both the resistivity and the AHE are very susceptible to a
small perpendicular field.
\par

Finally, we note that for the highest mobility state achieved  (blue stars), quantum oscillations are observed in both MR and Hall measurements [Fig.\ref{Hc_Tc}b, Fig.\ref{Rotation}b]. These oscillations will be analyzed separately.\cite{Quantumos}


\par
Spin-orbit interactions are expected to play a significant role at the interface
due to the inversion symmetry breaking.
The SO interaction term is of the form:
$\epsilon_{SO}=\mathbf{p}\cdot\mathbf{\sigma}\times\mathbf{\nabla}
V$ where $\mathbf{p}$ is the carrier momentum, $\mathbf{\sigma}$ is the Pauli spinor, and $\mathbf{\nabla}V$ is the electric field perpendicular to the interface in our system.\cite{Rashba}
The strong gate
dependence of \Eso~ directly follows from this expression. As pointed out above, the
initial local electric field is unknown. When
a positive gate voltage was applied for the first time the number of
charge carriers decreased, while the resistivity decreased. We believe
that this is mainly due to a decrease in the total electric field near
the interface, resulting in a decreased SO scattering and an enhanced mobility. This suggests that the initial field (AG state) is a consequence of the electronic reconstruction. In this scenario, adding more \LAO~ layers increases the as-grown field and hence the SO scattering. This explains the decrease in mobility while increasing the \LAO~ thickness.\cite{Thickness}
\par
It seems that most scattering processes in the normal-state involve a spin flip, which strongly
impede the transport. In a simple metal such processes do not contribute
much to the resistivity, yet in our system they become important due to the strong SO coupling.
This is clear in the case of in-plane spin flip processes; since SO coupling
affects in-plane spin orientations, a spin flip process results in
momentum reversal and consequently in a strong contribution to the
resistivity. These spins are
strongly coupled to the in-plane momenta; therefore it is impossible to align
them with \Hpar~ unless the energy associated with the field
$g\mu_B$\Hpar~exceeds $\epsilon_{SO}$.
For H$>$\Hstar, the spins gradually align along the field
direction and so R$_n$ is reduced. When all spins are aligned, the resistance reaches the
saturation value R$_{sat}$. For the out-of-plane field orientation, suppression of spin
scattering is overwhelmed by the large positive MR.
\par
Tuning the gate voltage from 50 to -50 V increases the
local electric field, resulting in an increase of $\epsilon_{SO}$
and \Hstar. Unlike R$_n$ (the zero field
resistance), R$_{sat}$ roughly scales with number of charge
carriers deduced from the high field Hall measurements.
This suggests that R$_{sat}$ is a result of standard impurity scattering.
For Vg=-50 V, $\epsilon_{SO}$ is extremely
large and an in-plane field of 18 T is not enough to align the
spins and suppress the spin-scattering resistivity.
Upon applying a small perpendicular field component and due to the
large scattering time, orbital motion is immediately turned
on along with the SO coupling. This results in a full
recovery of the spin-MR.
\par
In summary, we study the phase diagram of the
\STO$/$\LAO~interface in the region where T$_c$ increases, while
reducing the carrier concentration by variation of gate voltage.
We demonstrate the important effect of spin-orbit (SO) interaction
on both superconductivity and on normal-state transport. The SO
coupling energy (\Eso) is evaluated using two independent
transport properties, and is also in agreement with theoretical
model given the superconducting parameters: T$_c$ and the upper
critical parallel field. \Eso~ follows T$_c$ for the electric field range studied.
Our results suggest that \STO/\LAO~
interfaces may be useful for future oxide based devices
controlling the orbital motion of electrons by acting on their
spins. \cite{Wolf}
\par
We are indebted to G. Deutscher, A. Aharony, and Ora Entin-Wohlman for enlightening
discussions. This research was partially supported by the Bikura
program of the ISF grant No. 1543/08 by
the ISF grant No. 1421/08 and by the
Wolfson Family Charitable Trust. A portion of this work was
performed at the National High Magnetic Field Laboratory, which is
supported by NSF Cooperative Agreement No. DMR-0654118, by the
State of Florida, and by the DOE.

\bibliographystyle{apsrev}
\bibliography{spinorbit9}
\end{document}